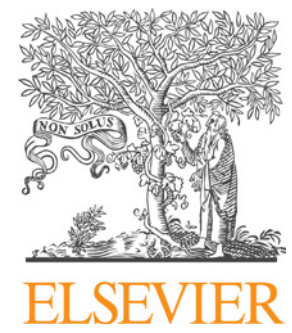

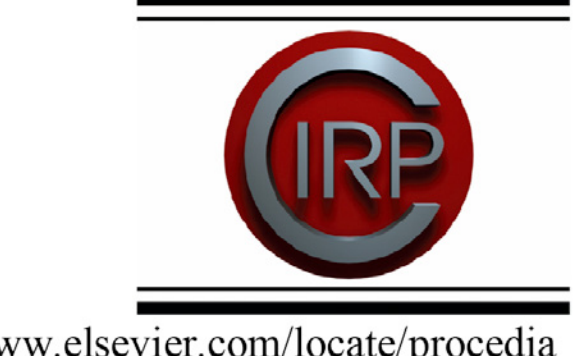

# IoT-Driven Cloud-based Energy and Environment Monitoring System for Manufacturing Industry

Nitol Saha[a,*,1] , Md Masruk Aulia[b,1], Md. Mostafizur Rahman[c,1], Abu Zahid[d], Mohammed Shafiul Alam Khan[e]

[a]University of South Carolina, Columbia, 29201, South Carolina, United States
[b]Military Institute of Science and Technology (MIST), Dhaka, 1216, Bangladesh
[c]Khulna University of Engineering & Technology, Khulna, 9203, Bangladesh
[d]Lamar University, Beaumont, 77705, Texas, United States
[e]Institute of Information Technology (IIT), University of Dhaka

* Corresponding author. Tel.: +1-803-269-8057; *E-mail address:* nsaha@email.sc.edu

**Abstract**

This research focused on the development of a cost-effective IoT solution for energy and environment monitoring geared towards manufacturing industries. The proposed system is developed using open-source software that can be easily deployed in any manufacturing environment. The system collects real-time temperature, humidity, and energy data from different devices running on different communication such as TCP/IP, Modbus, etc., and the data is transferred wirelessly using an MQTT client to a database working as a cloud storage solution. The collected data is then visualized and analyzed using a website running on a host machine working as a web client.

## 1. Introduction

In the fast-paced manufacturing environment of today, the pursuit of environmental sustainability and operational efficiency is more important than ever. To advance in this field, the incorporation of cloud computing and the Internet of Things (IoT) has emerged as a game-changer, presenting previously unheard-of possibilities for revolutionizing environmental monitoring and energy management in manufacturing facilities. The manufacturing sector is significantly responsible for the global energy consumption. Based on the periodic assessments and manual interventions, traditional approaches to energy management and environmental monitoring have been applied but the approaches could not comply with the complexities and dynamics of modern manufacturing processes. However, the convergence of IoT and cloud technologies presents a paradigm shift, enabling real-time, data-driven insights into energy usage, emissions, and environmental conditions within manufacturing facilities.

In the realm of smart manufacturing, data collection and monitoring stand as indispensable tools, enabling industries to harness the full potential of the Industrial Internet of Things (IIoT). With a strong focus on machine-to-machine (M2M) communication, IIoT is enabling a new era of smart factories, predictive maintenance, and data-driven decision-making. IoT protocols are purpose-built for machine-to-machine (M2M) communication. They minimize overhead, are highly responsive, and allow for efficient data transmission, even in low-bandwidth or unreliable network conditions. This makes them an excellent choice for industries that demand real-time data from numerous sensors and devices, ensuring seamless and reliable operations. Various communication protocols are currently utilized for IoT devices.

[1] Authors with equal contribution

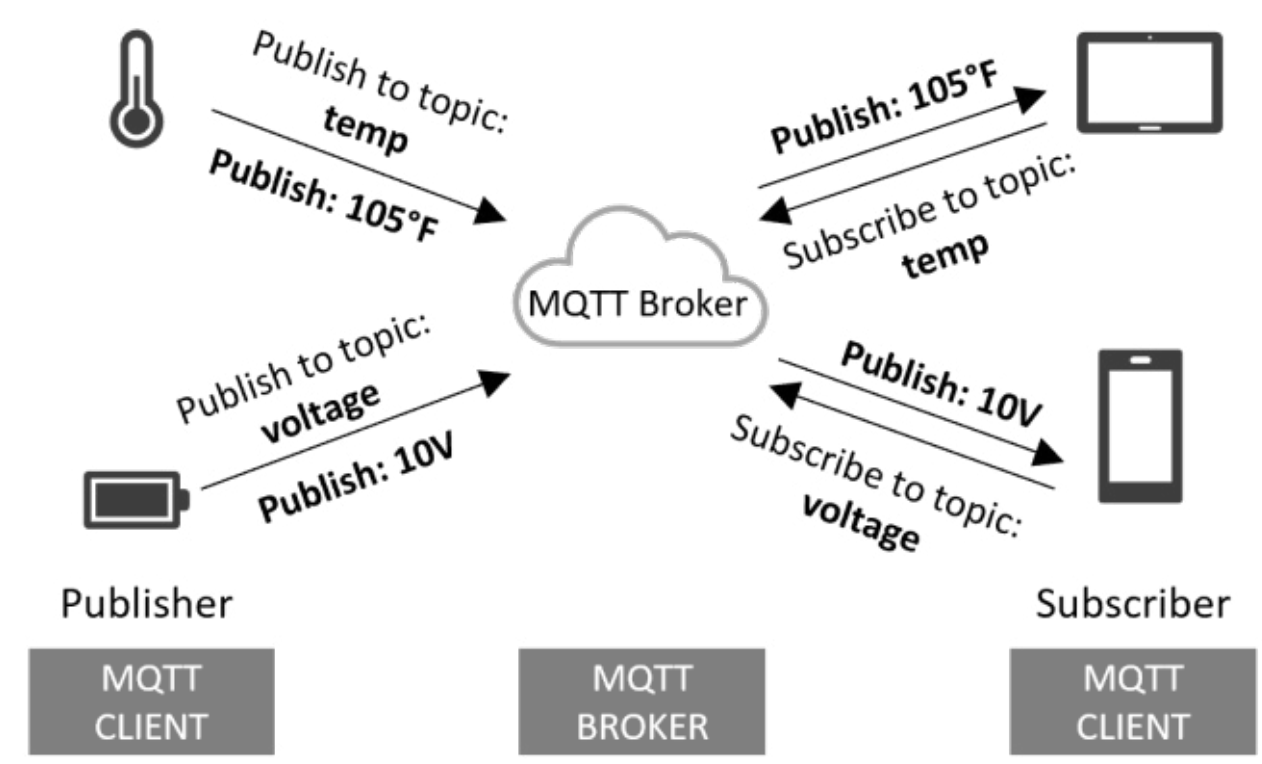

Fig. 1. MQTT Publish/Subscribe architecture

One of the IoT protocols is the MQTT protocol which is standardized by ISO (ISO / IEC 20922: 2016) [1]. Message Queuing Telemetry Transport (MQTT) protocol is a messaging protocol that uses a Publish/Subscribe mechanism (Fig. 1). It is currently in the OASIS (Organization for the Advancement of Structured Information Standards) standard. This protocol is used widely for IoT systems, and it is lightweight, has a small bandwidth requirement, and is easy to implement [2]. Commonly used communication protocols for industrial systems include Modbus, PROFINET, EtherNet/IP, DeviceNet, Fieldbus, Profibus, HART, CAN Bus, OPC, Modbus, and BACnet. These protocols facilitate data transfer, control, and real-time communication between devices, sensors, programmable logic controllers (PLCs), and control systems in industries.

This research work is aimed at implementing a low-cost IoT device and interface to common industrial communication protocols to collect and store versatile data on the cloud. This study explores the relevance, capabilities, and possible effects of IoT-Driven Cloud-based Energy and Environment Monitoring Systems on the manufacturing sector's pursuit of sustainable growth.

## 2. Related Work

To identify the most recent research works on the Energy and Environment Monitoring System, different pieces of literature were reviewed. Electricity monitoring consumption in buildings such as dormitories, hospitals, hotels, etc. was discussed in [3]. The IoT-inspired system is being used for each individual room of a building which shows users the complete statistical reports of their consumption. In [4], a decentralized metering architecture that allows IoT-capable devices to measure their own energy consumption has been proposed. The authors of [5] have proposed, designed, and implemented a cost-effective IoT-enabled three-phase smart energy meter that can be used to accumulate, process, and transmit information to any smart energy control system regarding electrical energy based on users. Specific software has been used to control the stored information and to perform data analysis and post-processing. The discussed system is a client-server-based and expandable system that has scopes of further integration. In paper [6], a cost-friendly IoT-based energy monitoring system has been implemented which is formulated on a PZEM-004T, using non-invasive CT sensors, SD3004 electric energy measurement chip, and ESP8266 Wemos D1 mini microcontroller for data that is retrieved from sensor nodes and sending data to server via the Internet. The data from the test run has demonstrated that the voltage, current, active, and accumulated power consumption can be recorded by the system. The proposed system in [7] accounts for the online display of the power usage of solar energy as renewable energy. Raspberry Pi along with the flask framework has been used to monitor the power usage. Daily usage of renewable energy is presented by smart monitoring. The paper [8] presents the design and implementation of an energy meter which is based on an Arduino microcontroller to measure the power consumed by any individual electrical appliance. It has an energy monitoring website that displays the calculated power consumption by various electrical devices. A real-time energy monitoring system based on IoT has been discussed in [9] which is used for controlling and monitoring the switchgear industry. Due to cost optimization and reliable technology, Raspberry Pi with Node.js programming language has been used to collect the electrical parameters from the existing energy meters and thus store them in a local drive so that the data can be accessed through a laptop or mobile device using Grafana. In order to understand the present-day energy pattern, the system has been found useful and it can certainly play a key role in energy conservation measures for minimizing energy consumption. Authors of [10] introduced a cost-effective IoT device utilizing PZEM-004t sensors, Arduino Nano Mini, and ESP8266 for monitoring electrical energy usage, showcasing successful functionality and potential for effective energy management. Research in [11] aims to develop an affordable IoT-based energy monitoring and management system, integrating smart device applications, cloud databases, and Raspberry Pi hardware for global control of electrical appliances to enhance energy efficiency.

Existing literature primarily focuses on energy monitoring in buildings and small-scale settings, with limited emphasis on manufacturing industries. The reviewed literature also lacks a comprehensive solution for integrating diverse devices with varying communication protocols for real-time data collection in industrial environments. This article addresses the gap by proposing a cost-effective IoT solution tailored for manufacturing industries utilizing open-source software for easy deployment and integration with devices employing different communication protocols. It facilitates real-time data collection of temperature, humidity, and energy, wirelessly transmitting it to a cloud-based storage solution for visualization and analysis.

## 3. System Design & Implementation

The system architecture of the proposed system encompasses several key components which are discussed below:

### *3.1. Industrial Energy Meter*

EasyLogic PM2100, an industrial energy meter, is utilized to monitor the energy consumption of a 3-phase industrial water pump. With Modbus support, the PM2100 facilitates the retrieval of various energy parameters for detailed analysis. Integration of PM2100 with the Schneider Electric Modbus Gateway Com'X enables seamless communication by getting the data using Modbus RS 485 protocol. This Modbus gateway is connected to a laptop that is working as a Modbus Client. RS485 to USB converter was used to retrieve the energy consumption data from the energy meter to the laptop. A Python script sends the data over MQTT. The energy meter is depicted in Fig. 2.

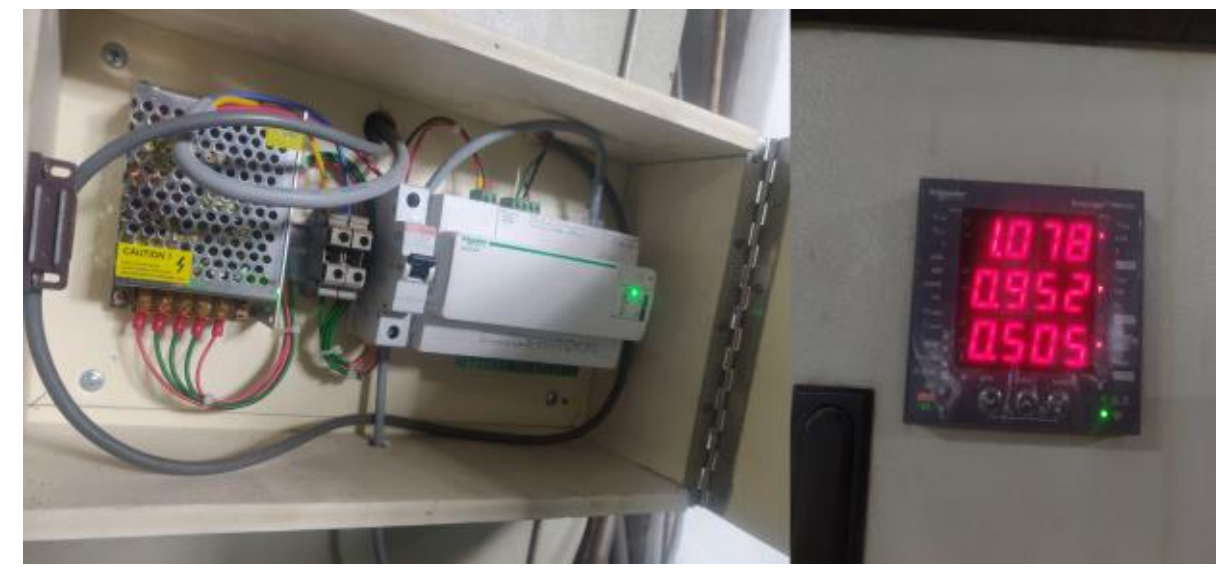

Fig. 2. Industrial energy meter

### *3.2. ESP32-based Energy Meter*

To create a low-cost energy meter that can be integrated with any electrical equipment, an ESP32-based energy meter, depicted in Fig. 3, is developed. Current Transformer (CT) module ZMCt103C and AC Voltage Transformer module ZMPT101B have been integrated with ESP32 to calculate the power using the code running on ESP32. This energy meter publishes the energy data over MQTT. The ESP32 collected analog values from CT and Voltage Transformer using analog ports [12].

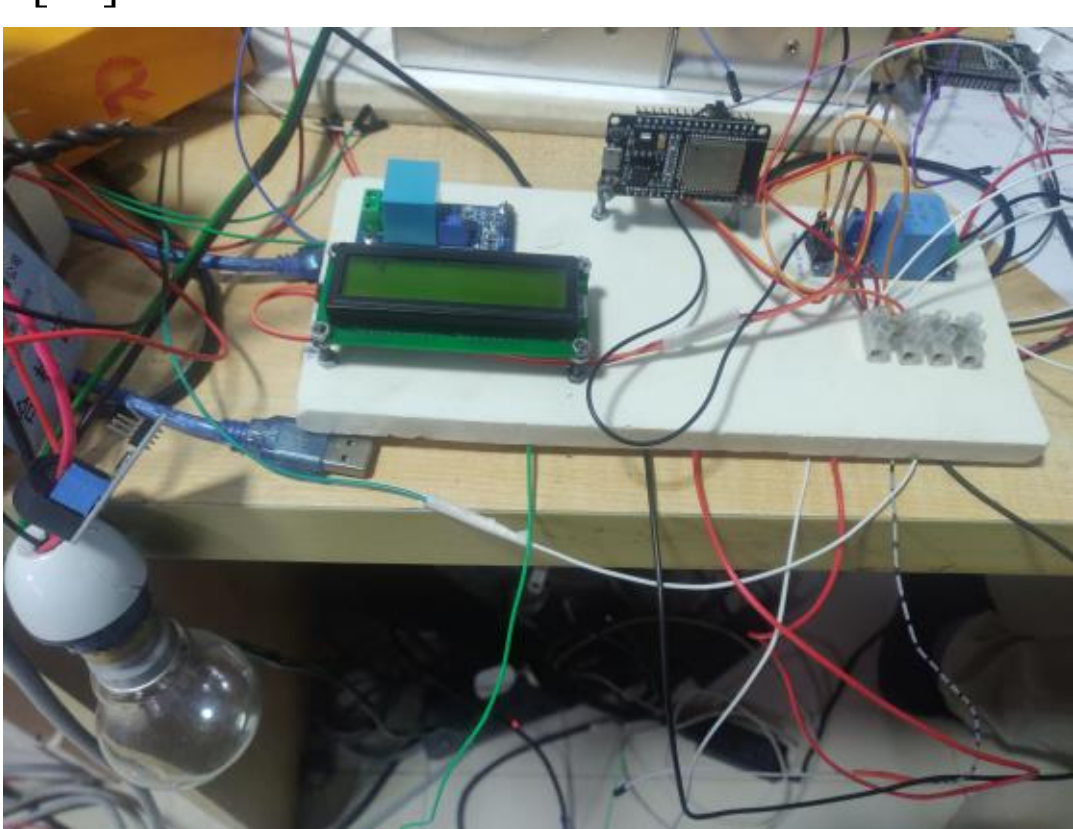

Fig. 3. ESP32-based Energy Meter

### *3.3. ESP32-based Environment Monitoring Device*

Temperature and Humidity composite sensor AM2303 is integrated with ESP32 to create an environment monitoring device that monitors the temperature and humidity of the environment (Fig. 4). This monitoring device publishes the temperature and humidity data over MQTT periodically [13].

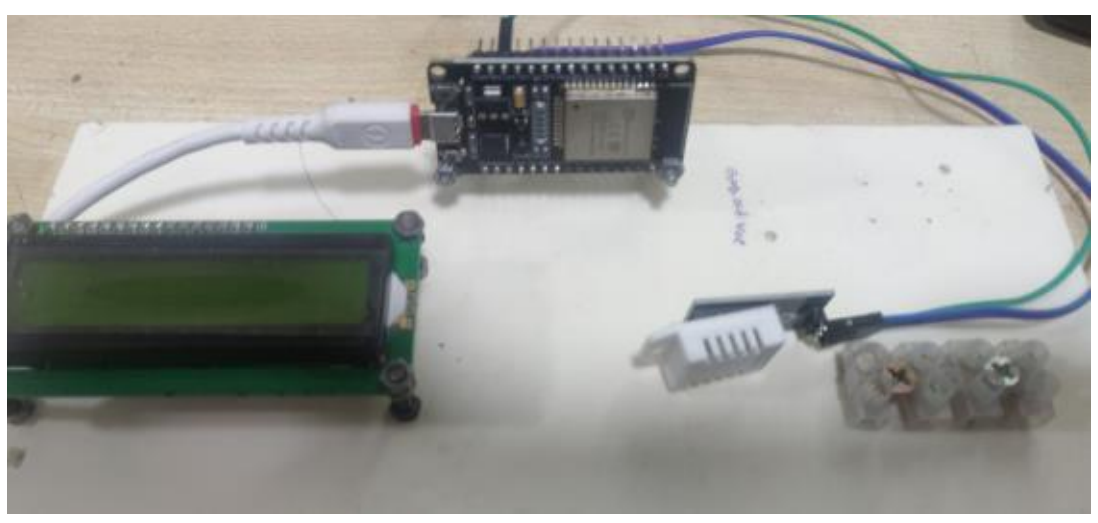

Fig. 4. ESP32-based Energy Meter

### *3.4. Raspberry-Pi-based MQTT Broker*

Raspberry Pi v4 was used as a local MQTT broker that processes the requests and responses from MQTT clients. Eclipse Mosquitto, an open source MQTT broker, is used as the MQTT broker. This works as an edge device in the manufacturing environment that processes all the local requests and responses in the plant and sends the data to the appropriate MQTT clients wirelessly for storing and real-time monitoring. This eliminates the need for multiple connections of the inter-connected sensors and works as a central hub to process multiple sensor data with bidirectional communication.

### *3.5. Web client*

The web client, functioning as an MQTT client, captures real-time data from sensors via the MQTT broker and stores this data within a MySQL server. Operating remotely on a laptop, it facilitates the visualization of this data through a web browser running on a local host. A dedicated website has been developed to monitor real-time data, with additional functionality allowing for the retrieval and visualization of historical data from the SQL server. This integrated system ensures seamless data management and visualization.

The system architecture of the proposed system is depicted in Fig. 5. The system comprises four MQTT clients and one MQTT broker. Three MQTT clients act as publishers, transmitting data from the industrial energy meter, ESP32-based energy monitoring device, and an environmental sensor to the MQTT broker. The fourth MQTT client serves as a subscriber, collecting data from the broker. The industrial energy meter values are shared using Modbus protocol to a local PC. This local PC publishes the energy meter's parameters periodically using MQTT protocols. The current and voltage sensors' values are periodically sent from the ESP32-based energy monitoring IoT device to the MQTT broker. The ESP32-based environment monitoring device collects temperature and relative humidity sensor data through I2C protocols and periodically sends it to the MQTT broker. All the data are published to the Raspberry Pi broker on three different topics. A Python program deployed on a remote PC subscribes to the MQTT topics to receive the data and it stores the data in the local database. An interactive website monitors trend data, hosted on a PHP server on the PC. AJAX was used to create the graphical plots of the data on the website.

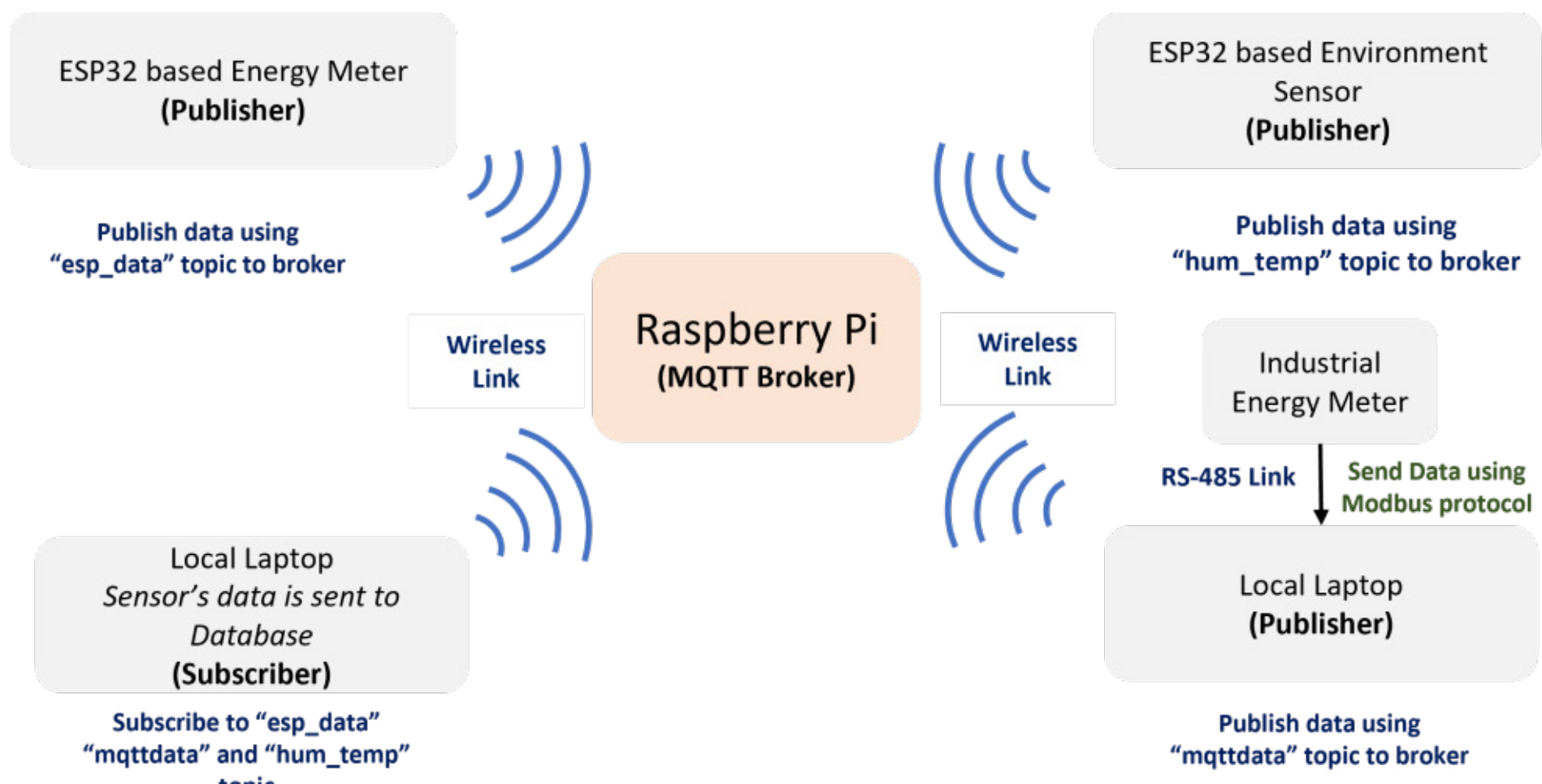

Fig. 5. System architecture of the proposed system

## 4. Results

The IoT-driven cloud-based energy and environment monitoring system for the manufacturing industry demonstrated robust performance across various metrics. The system's ability to collect, process, and analyze data in real-time significantly contributed to its effectiveness in monitoring energy consumption and environmental parameters within manufacturing facilities. When the system is started, IoT devices connected to the system start transmitting data using the MQTT protocol. The data is received by the subscriber's PC and stored in a database. The stored data can be viewed by using an internally developed webpage. There are several features on the webpage which are as follows:

### *4.1. Dashboard*

To access the system, a user has to log in using the user ID and password. After logging in, a dashboard is demonstrated containing various sections like a list of users, energy data, environment data, etc. (see Fig. 6)

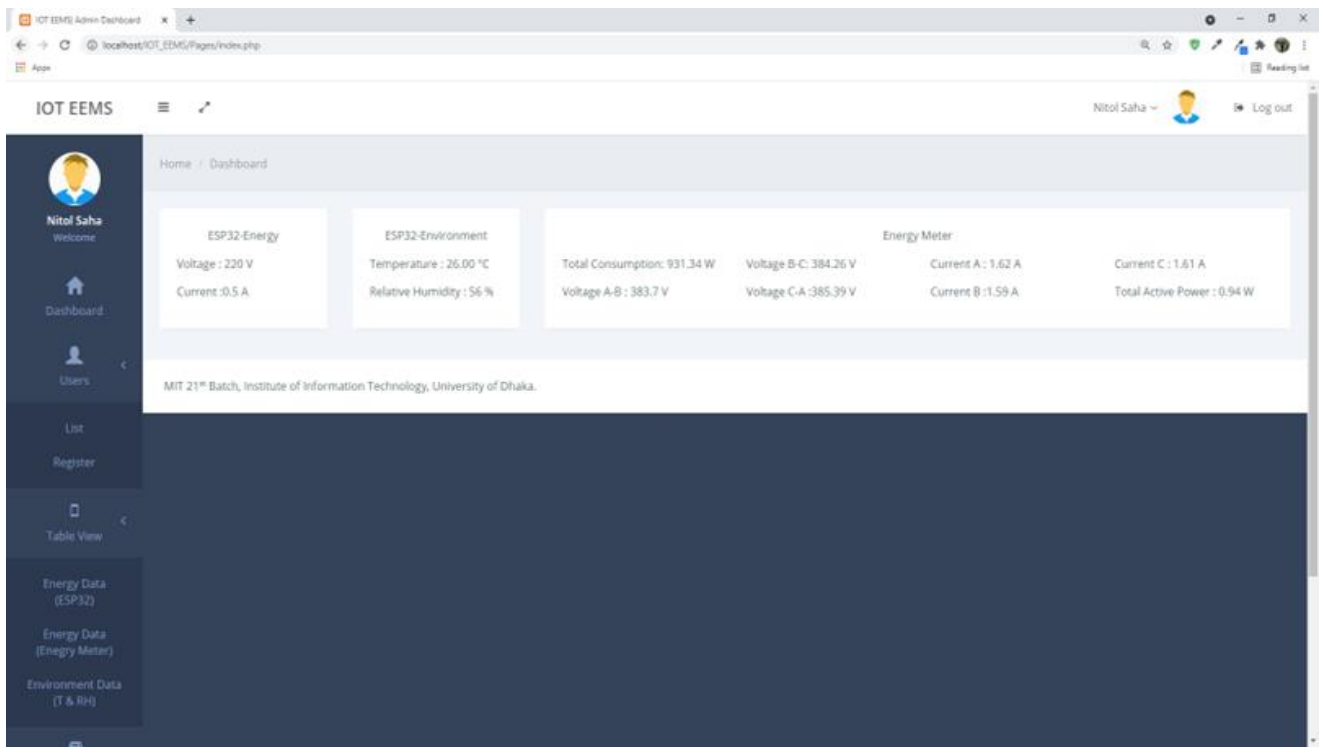

Fig. 6. Dashboard of the monitoring system

### *4.2. ESP32-based Energy Meter Data*

The energy consumption data collected from the EPS32 Energy IoT device can be viewed in tabular format depicted in Fig. 7. Data can be filtered according to a certain timeframe.

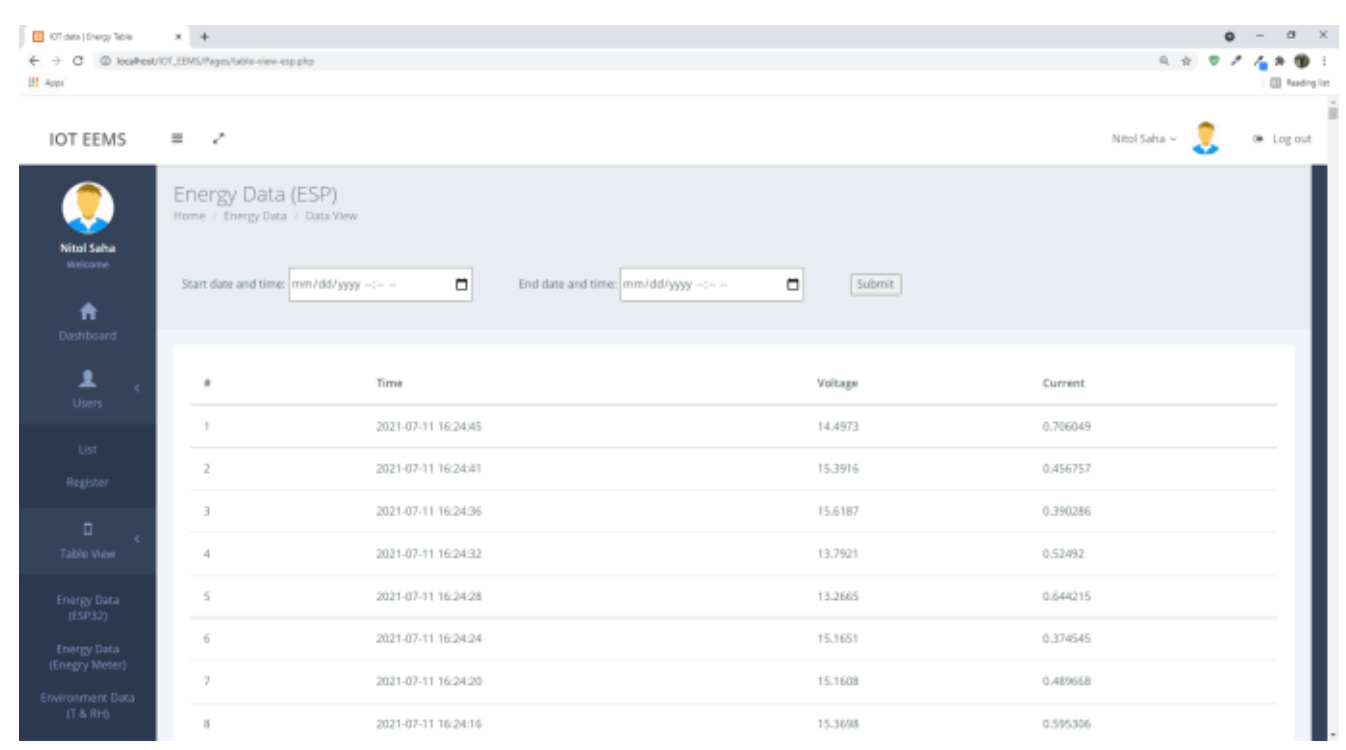

Fig. 7. ESP32-based energy meter data page

### *4.3. Industrial Energy Meter Data*

The energy consumption data (power) collected from the industrial energy meter data can also be viewed in tabular format (Fig. 8.). Data can be filtered according to a certain timeframe.

### *4.4. ESP32-based Environment Monitoring Device Data*

The system stores both temperature and humidity, which can be viewed in a tabular format shown in Fig. 9. Data can be filtered according to a certain timeframe.

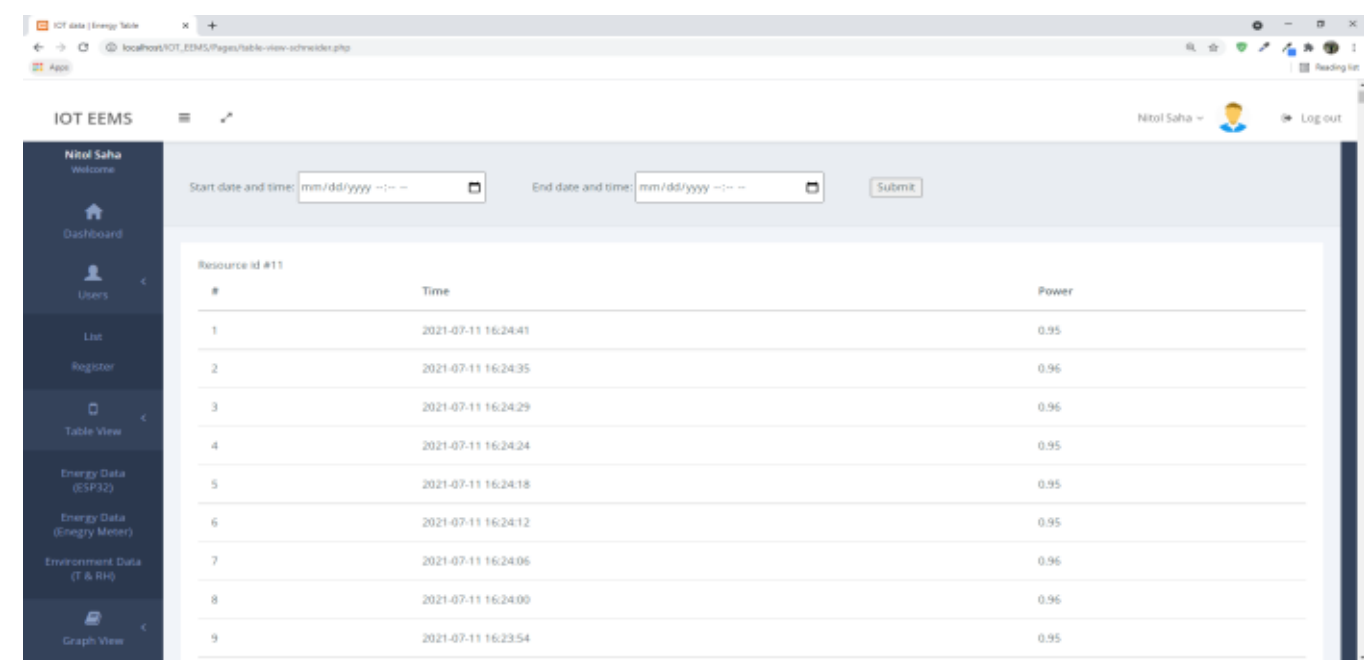

Fig. 8. Industrial energy meter data page

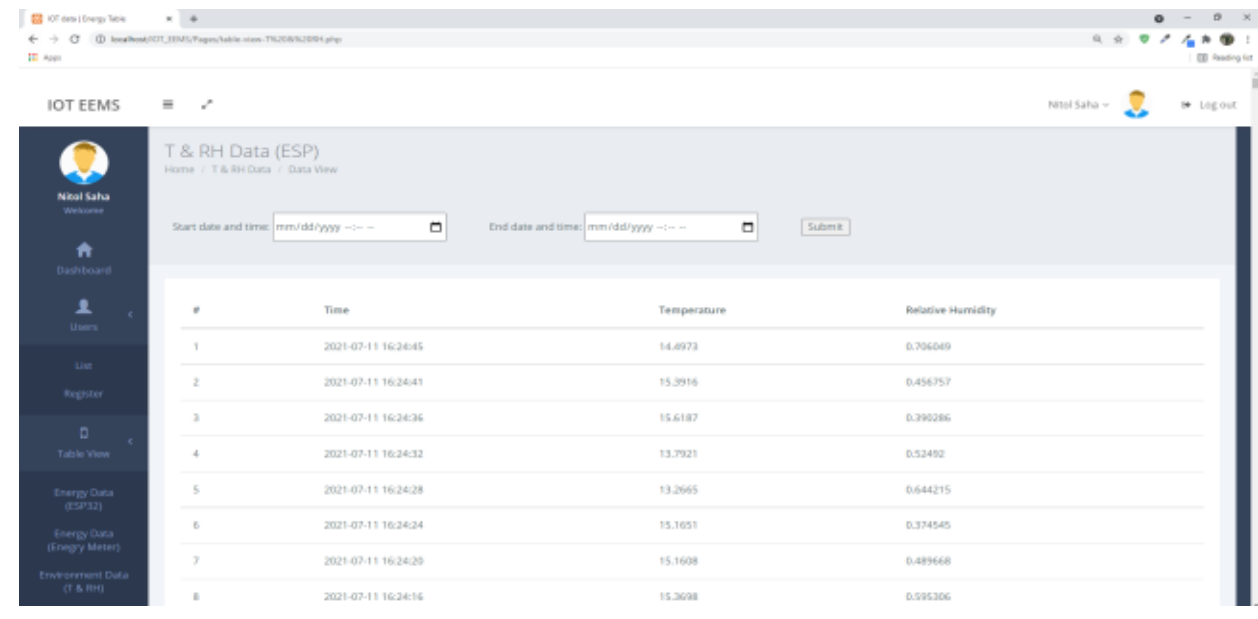

Fig. 9. Industrial energy meter data page

### *4.5. ESP32-based Energy Meter Plot*

The real-time current and voltage data from the EP32 Energy IoT device can also be viewed on the website depicted in Fig. 10. It shows trends in energy usage over time.

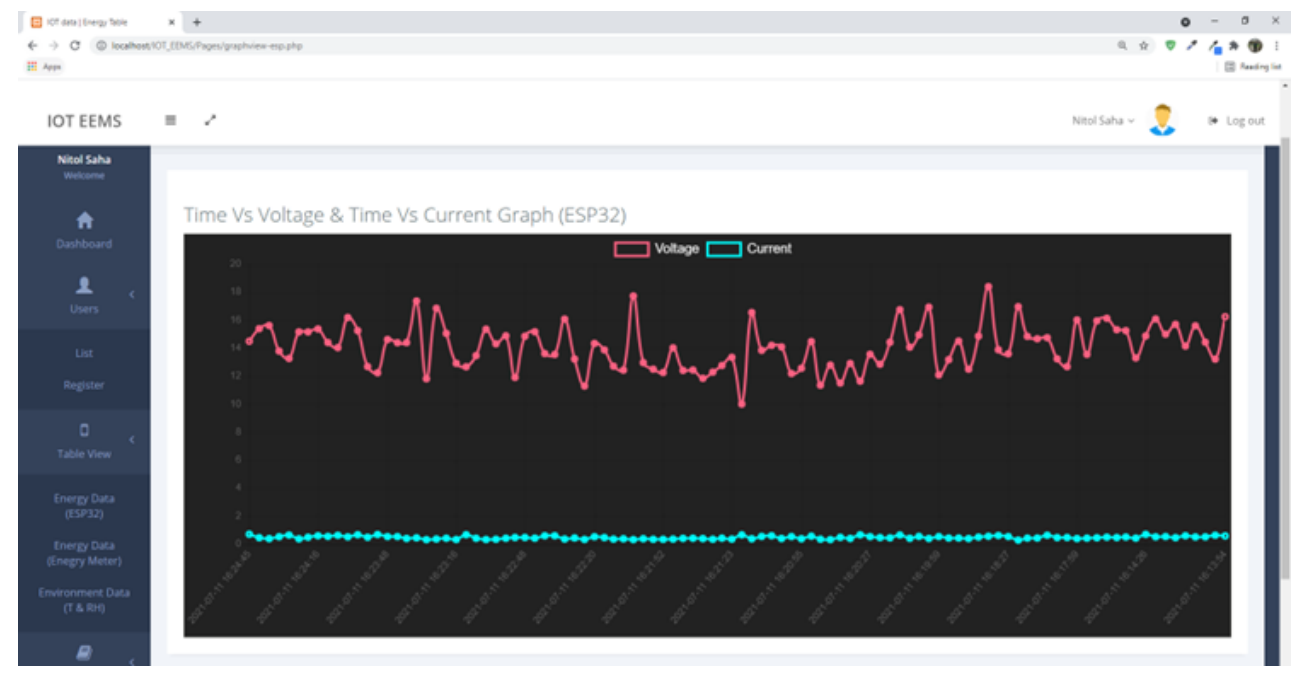

Fig. 10. The real-time plot of ESP32-based energy meter's data

### *4.6. Industrial Energy Meter Plot*

The real-time power data from the industrial energy meter data can be viewed using the website shown in Fig. 11. It shows trends in power usage over time.

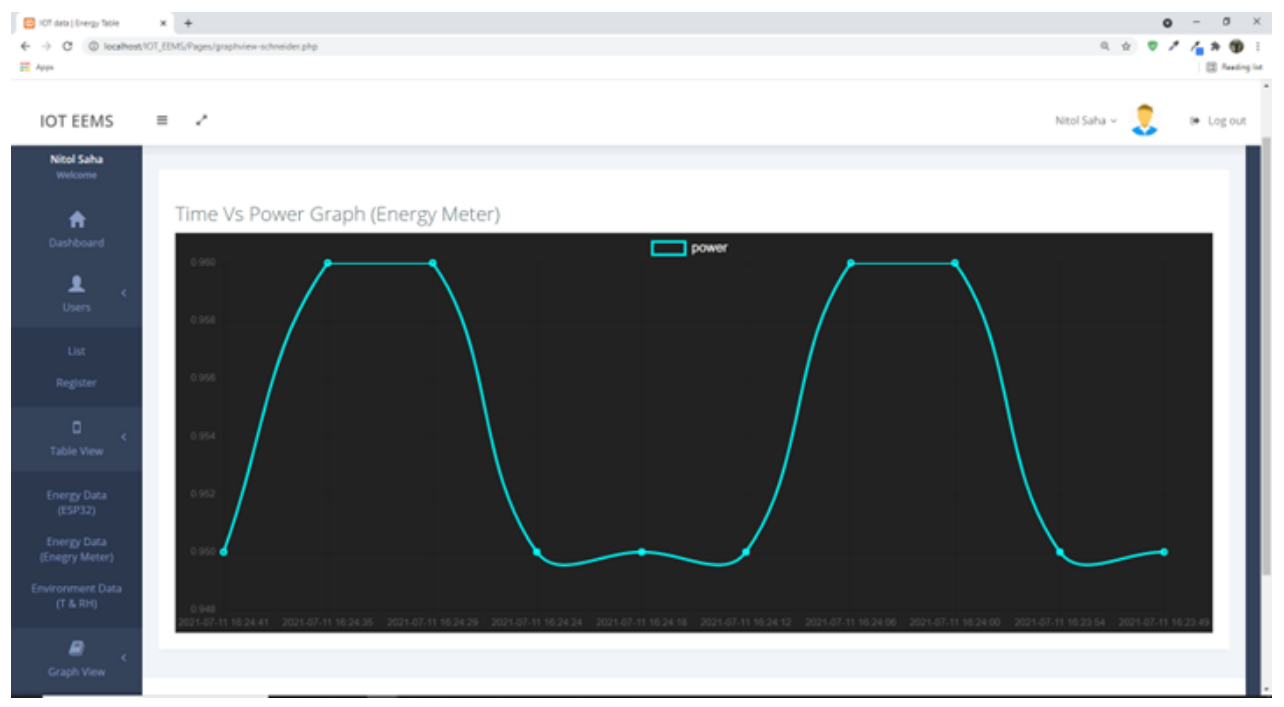

Fig. 11. The real-time plot of ESP32-based energy meter's data

### *4.7. ESP32-based Environment Monitoring Device Plot*

The real-time temperature and humidity data from the ESP32-based Environment Monitoring Device can be viewed using the website shown in Fig. 12. It shows trends of temperature and humidity data over time.

The IoT-driven cloud-based energy and environment monitoring system offers real-time tracking of energy usage and environmental conditions in manufacturing settings. It provides easy access to data through a user-friendly webpage, allowing users to analyze trends, monitor fluctuations, and make informed decisions. The system exhibited a high level of data accuracy and reliability, with minimal discrepancies observed between the measured and actual values of energy consumption and environmental parameters. Through rigorous testing and validation procedures, we ensured that sensor data were consistently precise and consistent, enabling reliable decision-making processes for facility managers. Its scalability and adaptability ensure compatibility with existing infrastructure.

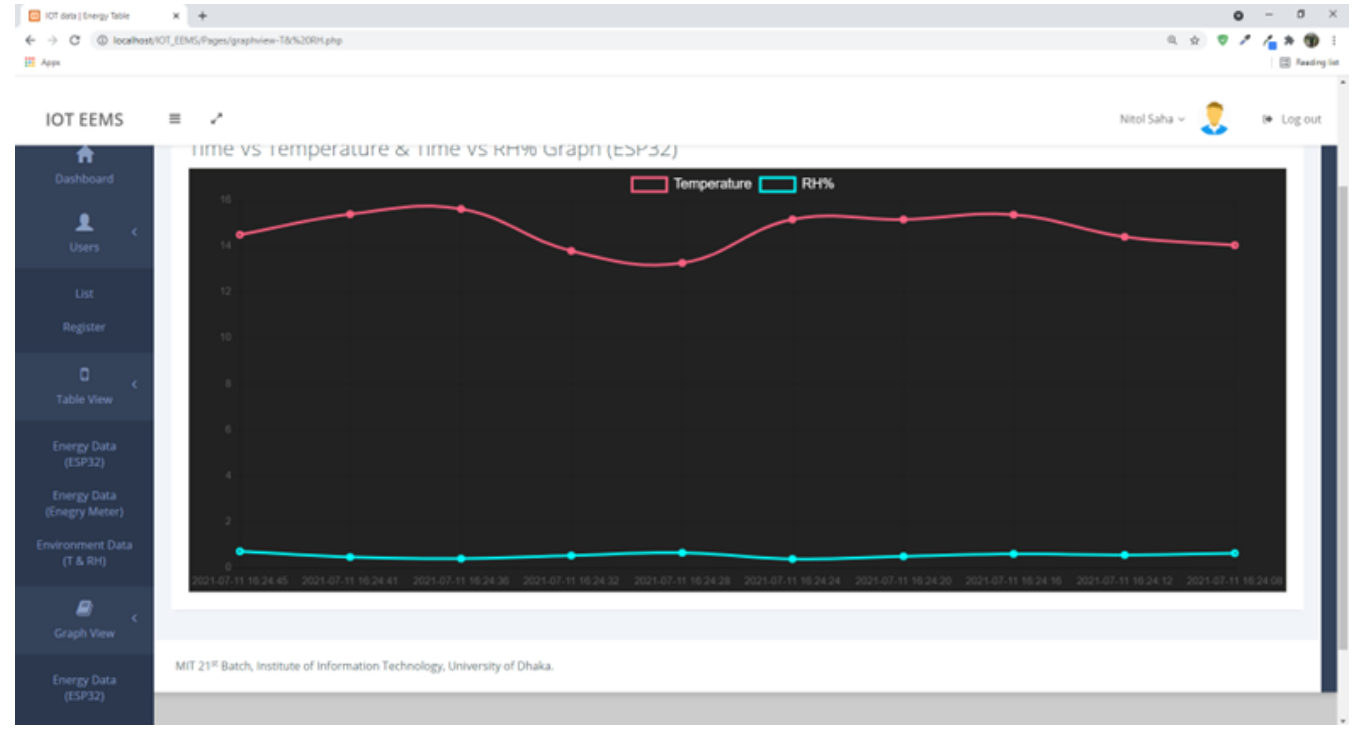

Fig. 12. The real-time plot of ESP32-based environment monitoring device's data

## 5. Conclusion

In this paper, a cloud-based environment and energy monitoring system has been proposed and discussed which consists of IoT devices to monitor and store energy data, temperature, and relative humidity. From the experimental results, it is evident that the IoT platform is successfully transmitting data to the remote database. The data can be easily accessed and demonstrated from the webpage in tabular format and also in live graphical representation. However, more research can be performed in this sector and perfection can only be achieved by continued research work. For further development of the system, integration of more sensors and protocols with logic will be performed. The application of machine learning in the system will also be implemented to analyze the stored data from the system and determine the predictive behavior of the devices so that synchronization among the devices can be increased.